\documentclass[twocolumn,english,aps]{revtex4-1}
\pdfoutput=1
\usepackage{ae,aecompl}
\usepackage[T1]{fontenc}
\usepackage[latin9]{inputenc}
\usepackage{geometry}
\usepackage{color}
\usepackage{babel}
\usepackage{amsmath}
\usepackage{amssymb}
\usepackage{graphicx}
\usepackage[unicode=true,pdfusetitle,
 bookmarks=true,bookmarksnumbered=false,bookmarksopen=false,
 breaklinks=false,pdfborder={0 0 1},backref=false,colorlinks=false]
 {hyperref}
\begin{document}
\title{A study of the nonlinear optical response of the plain graphene and
gapped graphene monolayers beyond the Dirac approximation}
\author{G. B. Ventura}
\email{corresponding author: gbventura@fc.up.pt}

\author{D. J. Passos}
\affiliation{Centro de F\'isica das Universidades do Minho e Porto}
\affiliation{Departamento de F\'isica e Astronomia, Faculdade de Ci\^encias,
Universidade do Porto, 4169-007 Porto, Portugal}
\author{J. M. Viana Parente Lopes}
\affiliation{Centro de F\'isica das Universidades do Minho e Porto}
\affiliation{Departamento de F\'isica e Astronomia, Faculdade de Ci\^encias,
Universidade do Porto, 4169-007 Porto, Portugal}
\author{J. M. B. Lopes dos Santos}
\affiliation{Centro de F\'isica das Universidades do Minho e Porto}
\affiliation{Departamento de F\'isica e Astronomia, Faculdade de Ci\^encias,
Universidade do Porto, 4169-007 Porto, Portugal}
\begin{abstract}
In this work, we present numerical results for the second and third
order conductivities of the plain graphene and gapped graphene monolayers
associated with the second and third harmonic generation, the optical
rectification and the optical Kerr effect. The frequencies considered
here range from the microwave to the ultraviolet portion of the spectrum,
the latter end of which had not yet been studied. These calculations
are performed in the velocity gauge \textcolor{black}{and directly
address the components of the conductivity tensor. In the velocity
gauge, the radiation field is represented by a power series in the
vector potential, and we discuss a very efficient way of calculating
its coefficients in the context of tight-binding models.}
\end{abstract}
\maketitle

\section{Introduction}

It has now been over a decade since the publication of the theoretical
works of S. A. Mikhailov on the low-frequency (\textit{intraband})
nonlinear response of the monolayer of graphene to an external electric
field \citep{mikhailov2007,mikhailov2008}, which marked the birth
of the study of nonlinear optical (NLO) responses in two-dimensional
materials. In the past ten years, this area has become increasingly
active and diverse, as it gathered the attention of both theoretical
\citep{Al-Naib2014,Glazov2014b,Peres2014,Cheng2014,Cheng2015b,Cheng2015,Mikhailov2016,semnani2016,Mikhailov2017,Savostianova:17,ventura,passos,Savostianova2018,Hipolito2018}
and experimental groups \citep{Hendry2010,Dean2010,Zhang2012,Kumar2013,Hong2013,Dremetsika2016,Vermeulen2016,Higuchi2017,Baudisch2018}.
This has also been extended to many other, more recently isolated,
layered materials \citep{Janisch2014,Pedersen2015,Hipolito2016,Hipolito2017,Youngblood2017}.
In those materials, like in graphene, the nonlinear optical response
has been shown to be very intense, much more so than in three dimensional
materials.

One key issue, that followed directly from those initial works, was
to expand the understanding of the nonlinear \textit{intraband} response
\textemdash{} frequencies in the microwave and the infrared \textemdash{}
into the high frequency range \textemdash{} frequencies in the near
infrared and above \citep{Cheng2014,Cheng2015b,Cheng2015,Mikhailov2016}.
Doing so required a full quantum treatment of the electrons in a crystal,
and meant recovering the formalism for the calculation of NLO coefficients
in bulk semiconductors of the late eighties and early nineties, developed
by J. E. Sipe and collaborators \citep{Moss1989,Sipe1993,Aversa1995,Sipe2000}.
\textcolor{black}{Their work, mostly formulated in the so-called length
gauge, provided expressions for the second and third order optical
conductivities that are directly applicable to a system of non-interacting
electrons in a solid, taking both intraband and interband transitions
into account. Many other works have since used this framework. In
practice, due to the complexity of the general expressions, calculations
of nonlinear optical conductivities usually require performing the
analytical calculation (i.e., an integration over the FBZ) for the
particular system under study: in third order this is already rather
cumbersome. Often, it is only really tractable for simple effective
Hamiltonians (such as the Dirac Hamiltonian in graphene), that describe
only a portion of the FBZ. This has limited the length gauge method
to sufficiently small frequencies for such effective Hamiltonians
to be applicable.}

\textcolor{black}{Another approach, based on the velocity gauge, was
developed concurrently but presented early difficulties. Spurious
divergences and inaccurate results upon the truncation of the number
of bands led the velocity gauge to be less adopted. The origin of
these difficulties was understood early on as a violation of sum rules
\citep{Aversa1995}. This was solved only recently \citep{passos},
with a reformulation of the velocity gauge that is able to reproduce
the results from the length gauge and that is best suited for numerical
calculations that involve the full FBZ. Two diagrammatic methods based
on this formulation of the velocity gauge have since been developed
\citep{parker2019,Joao2018}, the former of which was used in the
study of Weyl semimetals, while the latter was shown to be applicable
even in disordered systems. In this velocity gauge approach there
is no added dificulty in moving to higher frequencies and in fact
its implementation requires the use of models defined in the entire
FBZ. The authors will use the new velocity gauge approach of ref.\citep{passos}
to probe the NLO response of graphene in a frequency range beyond
the Dirac approximation.}

We present numerical results for the second and third order responses
of the plain graphene (PG) and gapped (GG) graphene monolayers to
a monochromatic electric field of frequencies (energies) that range
from the microwave ($\hbar\omega\sim0.005$ eV) to the ultraviolet
($\hbar\omega\sim6$ eV). These results differ from what has been
previously reported in literature \citep{Cheng2014,Cheng2015,Cheng2015b,Mikhailov2016,passos,Hipolito2018}
for two reasons: \textcolor{black}{we go beyond the Dirac cone approximation
(valid up to about 1 eV) and study the response of the PG and GG monolayers
at high frequencies.} Our calculations address all different components
of the conductivity tensors \textemdash{} on which intrinsic permutation
symmetry is imposed \textemdash{} and not the \textit{effective} tensors
of ref.\citep{Hipolito2018}, which also goes beyond the Dirac approximation,
but where Kleinman's symmetry is additionally imposed. Since this
second symmetry follows from the consideration that the nonlinear
susceptibilities (or conductivities) can be deemed \textit{dispersionless}
\citep{Boyd:2008}, it lacks justification in the study of the response
in these frequency ranges. Although seemingly technical in nature,
this difference is practically relevant as the conductivities computed
here can be directly related to measurements of the current response,
$J^{\alpha}(t)$, in an experiment (regardless of the polarization
of the electric field), whereas the effective tensors cannot.

The paper is organized as follows. In the following section, we perform
a review of the calculations of NLO conductivities in the length and
the velocity gauges. Section \ref{sec:III} is dedicated to the use
of tight-binding Hamiltonians in velocity gauge calculations and to
two pertinent points: the computation of $h$ coefficients, which
are integral to the description of the response in the velocity gauge
become simple when working on a basis for which the Berry connections
are all trivial; the second point concerns the relation between these
Berry connections and the manner by which one defines the position
operator in the lattice. It is shown that this has implications in
the optical response by studying the interband portion of the linear
conductivity of plain graphene. In Section \ref{sec:IV}, we present
the aforementioned results, i.e., the second harmonic generation and
optical rectification conductivities, of the GG monolayer and for
the third order response, i.e., the harmonic generation and Kerr effect
conductivities, of the PG and GG monolayers, in the aforementioned
frequency regime. For the two second order effects the results are
complemented by analytical calculations of the real part of the conductivities.
The final section is dedicated to a summary of our work.

\section{Calculation of nonlinear optical conductivities in crystals\label{sec:II}}

A system's nonlinear current response to a monochromatic electric
field, that is considered to be constant throughout the material,
\begin{align}
\mathbf{E}(t)= & \mathbf{E}_{0}\,e^{i\omega t}+\left(\mathbf{E}_{0}\right)^{*}\,e^{-i\omega t}
\end{align}
is described, in second order, by the second harmonic generation,
$\sigma_{\beta\alpha_{1}\alpha_{2}}^{(2)}(\omega,\omega)$, and the
optical rectification, $\sigma_{\beta\alpha_{1}\alpha_{2}}^{(2)}(\omega,-\omega)$,
conductivities,\footnote{We consider two things in the following expressions: repeated cartesian
indexes are being implicitly summed over; conductivities satisfy the
property of intrinsic permutation symmetry \citep{Boyd:2008}.}
\begin{align}
J_{\beta}^{(2)}(t)= & \ \sigma_{\beta\alpha_{1}\alpha_{2}}(\omega,\omega)\,E_{0}^{\alpha_{1}}\,E_{0}^{\alpha_{2}}\,e^{-i2\omega t}\nonumber \\
 & +\sigma_{\beta\alpha_{1}\alpha_{2}}(\omega,-\omega)\,E_{0}^{\alpha_{1}}\,(E_{0}^{\alpha_{2}})^{*}\nonumber \\
 & +\text{c.c.},
\end{align}
while, in third order, it is described by the third harmonic generation,
$\sigma_{\beta\alpha_{1}\alpha_{2}\alpha_{3}}^{(2)}(\omega,\omega,\omega)$
and the Kerr effect, $\sigma_{\beta\alpha_{1}\alpha_{2}\alpha_{3}}^{(3)}(\omega,\omega,-\omega)$,
conductivities,
\begin{align}
J_{\beta}^{(3)}(t)= & \ \sigma_{\beta\alpha_{1}\alpha_{2}\alpha_{3}}(\omega,\omega,\omega)\,E_{0}^{\alpha_{1}}\,E_{0}^{\alpha_{2}}\,E_{0}^{\alpha_{3}}\,e^{-i3\omega t}\nonumber \\
 & +\sigma_{\beta\alpha_{1}\alpha_{2}\alpha_{3}}(\omega,\omega,-\omega)\,E_{0}^{\alpha_{1}}\,E_{0}^{\alpha_{2}}\,(E_{0}^{\alpha_{3}})^{*}\,e^{-i\omega t}\nonumber \\
 & +\text{c.c.},
\end{align}
 The problem of studying $J_{\alpha}^{(n)}(t)$ is thus a problem
of knowing how to calculate the conductivities, $\sigma^{(n)}$, by
means of a perturbative expansion. This topic has been the subject
of intense research for crystalline systems \citep{Moss1989,Sipe1993,Aversa1995,Sipe2000,Cheng2014,Mikhailov2016,ventura,passos}
and we will use, in particular, results of our previous work \citep{ventura,passos},
in the following review of those calculations. The baseline considerations
here are the same as before: the electric field is constant throughout
the crystal and electron-electron interactions, integral to a description
of an excitonic response, are not taken into account.

\subsection{Crystal hamiltonian and its perturbations}

In a perfect infinite crystal, the eigenfunctions of the unperturbed
(crystal) Hamiltonian, $H_{0}$, are, according to Bloch's theorem,
written in terms of a plane wave and a function that is periodic in
the real space unit cell, 
\begin{align}
\psi_{\mathbf{k}s}(\mathbf{r})= & \ e^{i\mathbf{k}\cdot\mathbf{r}}\,u_{\mathbf{k}s}(\mathbf{r}),\label{eq:BLOCH_STATES}
\end{align}
for $\mathbf{R}$, any lattice vector,
\begin{align}
u_{\mathbf{k}s}(\mathbf{r})= & u_{\mathbf{k}s}(\mathbf{r}+\mathbf{R}).\label{eq:PERIODIC_STATES}
\end{align}
Each of these eigenfunctions and its corresponding eigenvalue, $\epsilon_{\mathbf{k}s}$,
is labelled by a crystal momentum, $\mathbf{k}$, that runs continuously
throughout the first Brillouin zone (FBZ) and by the $s$ index, indicating
the band. For a $d$ dimensional crystal, their normalization reads,
\begin{align}
\left\langle \psi_{\mathbf{k}s}\right|\left.\psi_{\mathbf{k}'s'}\right\rangle = & (2\pi)^{d}\,\delta_{ss'}\,\delta(\mathbf{k}-\mathbf{k}').
\end{align}
Furthermore, the periodic part of the Bloch functions (for a fixed
$\mathbf{k}$) also forms an orthogonal basis, with an inner product
that is defined over the real space unit cell, instead of the entire
crystal,
\begin{align}
\left\langle u_{\mathbf{k}s}\right|\left.u_{\mathbf{k}s'}\right\rangle = & \frac{1}{v_{c}}\int_{v_{c}}d^{3}\mathbf{r}\,u_{\mathbf{k}s}^{*}(\mathbf{r})\,u_{\mathbf{k}s'}(\mathbf{r})=\delta_{ss'}.\label{eq:SCA_P}
\end{align}
One can then write the full Hamiltonian, composed of the crystal Hamiltonian
and the coupling of the electrons to the external electric field,
in the single particle basis of band states. The explicit form of
the coupling depends on the representation one chooses for the electric
field in terms of the scalar ($\phi(\mathbf{r},t)$) and vector potential
($\mathbf{A}(\mathbf{r},t)$), i.e., on the chosen gauge.

For the length gauge, the vector potential is set to zero,
\begin{equation}
\mathbf{E}(t)=-\nabla\phi(\mathbf{r},t),
\end{equation}
and the coupling to the electrons is performed via dipole interaction,
$V_{\mathbf{k}ss'}^{E}(t)$,
\begin{align}
H^{E}= & \ \int\frac{d^{d}\mathbf{k}}{(2\pi)^{d}}\sum_{s,s'}\left|\psi_{\mathbf{k}s}\right\rangle \bigl[\epsilon_{\mathbf{k}s}\,\delta_{ss'}+V_{\mathbf{k}ss'}^{E}(t)\bigr]\left\langle \psi_{\mathbf{k}s'}\right|,\label{eq:LENGTH_H}
\end{align}
where,
\begin{align}
V_{\mathbf{k}ss'}^{E}(t)= & \ ie\mathbf{E}(t)\cdot\mathbf{D}_{\mathbf{k}ss'}.\label{eq:PERT_E}
\end{align}
The covariant derivative, $\mathbf{D}_{\mathbf{k}ss'}$, is defined
as \citep{ventura},
\begin{align}
\mathbf{D}_{\mathbf{k}ss'}= & \ \nabla_{\mathbf{k}}\delta_{ss'}-i\boldsymbol{\xi}_{\mathbf{k}ss'},\label{eq:COV_DEV}
\end{align}
for $\boldsymbol{\xi}_{\mathbf{k}ss'}$, the Berry connection between
band states \citep{Blount1962.},
\begin{align}
\boldsymbol{\xi}_{\mathbf{k}ss'}= & \ i\left\langle u_{\mathbf{k}s}\right|\left.\nabla_{\mathbf{k}}u_{\mathbf{k}s'}\right\rangle .
\end{align}
As for the velocity gauge, it is the scalar potential that is set
to zero,
\begin{align}
\mathbf{E}(t)= & -\partial_{t}\mathbf{A}(t),
\end{align}
and the full Hamiltonian in this gauge, $H^{A}$, is obtained from
$H^{E}$ by means of a time-dependent unitary transformation \citep{passos,ventura},
\begin{align}
H^{A}= & \int\frac{d^{d}\mathbf{k}}{(2\pi)^{d}}\sum_{s,s'}\left|\psi_{\mathbf{k}s}\right\rangle \bigl[\epsilon_{\mathbf{k}s}\,\delta_{ss'}+V_{\mathbf{k}ss'}^{A}(t)\bigr]\left\langle \psi_{\mathbf{k}s'}\right|,\label{eq:VELOCITY_H}
\end{align}
for a perturbation, $V_{\mathbf{k}ss'}^{A}(t)$, that is written as
an infinite series in the external field, 
\begin{align}
V_{\mathbf{k}ss'}^{A}(t)= & \sum_{n=1}^{\infty}\frac{e^{n}}{n!}\,A_{\alpha_{1}}(t)(...)A_{\alpha_{n}}(t)\,h_{\mathbf{k}ss'}^{\alpha_{1}(...)\alpha_{n}}.\label{eq:PERT_A}
\end{align}
The coefficients in that expansion, $h_{\mathbf{k}ss'}^{\alpha_{1}(...)\alpha_{n}}$,
are given by nested commutators of the covariant derivative, Eq.(\ref{eq:COV_DEV}),
with the unperturbed Hamiltonian \citep{passos}, 
\begin{align}
h_{\mathbf{k}ss'}^{\alpha_{1}(...)\alpha_{n}}= & \left\langle u_{\mathbf{k}s}\right|(\nabla_{\mathbf{k}}^{\alpha_{n}}...\nabla_{\mathbf{k}}^{\alpha_{1}}H_{0\mathbf{k}})\left|u_{\mathbf{k}s'}\right\rangle ,\label{eq:AITCH}\\
= & \frac{1}{\hbar^{n}}\bigl[D_{\mathbf{k}}^{\alpha_{n}},\bigl[(...),\bigl[D_{\mathbf{k}}^{\alpha_{1}},\,H_{0}\bigr]\bigr]\,(...)\bigr]_{ss'},
\end{align}
with the first one being the velocity matrix element in the unperturbed
system. Finally, one can write the velocity operator in each of the
gauges: $v^{\beta}=\hbar^{-1}\left[D^{\beta},H\right]$. In the single
particle basis, they read as
\begin{align}
v^{E,\beta}= & \int\frac{d^{d}\mathbf{k}}{(2\pi)^{d}}\sum_{s,s'}\left|\psi_{\mathbf{k}s}\right\rangle v_{\mathbf{k}ss'}^{(0),\beta}\left\langle \psi_{\mathbf{k}s'}\right|,\label{eq:VEL_NOT}\\
v^{A,\beta}(t)= & \int\frac{d^{d}\mathbf{k}}{(2\pi)^{d}}\sum_{s,s'}\left|\psi_{\mathbf{k}s}\right\rangle \Bigl[v_{\mathbf{k}ss'}^{(0),\beta}+\sum_{n=1}^{\infty}\frac{e^{n}}{n!}\,\nonumber \\
 & \times A_{\alpha_{1}}(t)(...)A_{\alpha_{n}}(t)\,h_{\mathbf{k}ss'}^{\beta\alpha_{1}(...)\alpha_{n}}\Bigr]\left\langle \psi_{\mathbf{k}s'}\right|.\label{eq:VEL_A}
\end{align}

\subsection{Density matrix and conductivities}

The electric current density in the crystal is given by the ensemble
average of the velocity operator times the charge of an electron,
\begin{align}
\langle J^{\beta}\rangle(t)= & \ (-e)\,\text{Tr}\left[v^{\beta}\rho(t)\right],\\
= & \ (-e)\int\frac{d^{d}\mathbf{k}}{(2\pi)^{d}}\sum_{s,s'}v_{\mathbf{k}s's}^{\beta}\,\rho_{\mathbf{k}ss'}(t),\label{eq:CURRENT}
\end{align}
and it is written in terms of the matrix elements of the density matrix
(DM), whose time evolution is described by the Liouville equation,
\begin{align}
(i\hbar\partial_{t}-\Delta\epsilon_{\mathbf{k}ss'})\rho_{\mathbf{k}ss'}(t)= & \left[V_{\mathbf{k}},\,\rho_{\mathbf{k}}(t)\right]_{ss'}.\label{eq:EQM}
\end{align}
Each gauge has its own set of equations of motion, following from
the perturbations of Eqs.(\ref{eq:PERT_E}) and (\ref{eq:PERT_A}).
The perturbative treatment of the current response requires an expansion
of the $\rho_{\mathbf{k}ss'}(t)$ in powers of the electric field
and solving \textemdash{} recursively \textemdash{} the equations
of motion for the matrix elements of the DM, Eq.(\ref{eq:EQM}), in
frequency space. For the velocity gauge, it also requires an expansion
of the velocity matrix elements, Eq.(\ref{eq:VEL_A}), since these
also depend on the electric field. At the end of that procedure \citep{passos,ventura},
one obtains the conductivities of arbitrary order $n$ in both the
length and velocity gauges. Here we only present the expressions for
the second order conductivities, following the scattering prescription
described in \citep{passos},
\begin{align}
\sigma_{\beta\alpha_{1}\alpha_{2}}^{(2),E}(\omega_{1},\omega_{2})= & e^{3}\int\frac{d^{d}\mathbf{k}}{(2\pi)^{d}}\sum_{s,s'}\frac{h_{\mathbf{k}s's}^{\beta}}{\hbar\omega_{12}+2i\gamma-\Delta\epsilon_{\mathbf{k}ss'}}\nonumber \\
 & \ \times\bigl[D_{\mathbf{k}}^{\alpha_{1}},\frac{1}{\hbar\omega_{2}+i\gamma-\Delta\epsilon}\circ\bigl[D_{\mathbf{k}}^{\alpha_{2}},\rho_{\mathbf{k}}^{(0)}\bigr]\bigr]_{\mathbf{k}ss'}\nonumber \\
 & +(\{\alpha_{1},\omega_{1}\}\leftrightarrow\{\alpha_{2},\omega_{2}\}),\label{eq:SIGMA_E}
\end{align}

\begin{flushleft}
\begin{widetext}
\begin{align}
\sigma_{\beta\alpha_{1}\alpha_{2}}^{(2),A}(\omega_{1},\omega_{2})= & \frac{e^{3}}{\omega_{1}\omega_{2}}\int\frac{d^{d}\mathbf{k}}{(2\pi)^{d}}\sum_{s,s'}\bigl[\frac{h_{\mathbf{k}s's}^{\beta}}{\hbar\omega_{12}+2i\gamma-\Delta\epsilon_{\mathbf{k}ss'}}\bigl(\bigl[h_{\mathbf{k}}^{\alpha_{1}},\frac{1}{\hbar\omega_{2}+i\gamma-\Delta\epsilon}\circ\bigl[h_{\mathbf{k}}^{\alpha_{2}},\rho_{\mathbf{k}}^{(0)}\bigr]\bigr]_{\mathbf{k}ss'}+\frac{1}{2}\bigl[h_{\mathbf{k}}^{\alpha_{1}\alpha_{2}},\rho_{\mathbf{k}}^{(0)}\bigr]_{\mathbf{k}ss'}\bigr)\nonumber \\
 & +h_{\mathbf{k}s's}^{\beta\alpha_{1}}\,\frac{1}{\hbar\omega_{2}+i\gamma-\Delta\epsilon_{\mathbf{k}ss'}}\bigl[h_{\mathbf{k}}^{\alpha_{2}},\rho_{\mathbf{k}}^{(0)}\bigr]_{\mathbf{k}ss'}+\frac{1}{2}h_{\mathbf{k}s's}^{\beta\alpha_{1}\alpha_{2}}\,\rho_{\mathbf{k}ss'}^{(0)}\ +(\{\alpha_{1},\omega_{1}\}\leftrightarrow\{\alpha_{2},\omega_{2}\})\bigr].
\end{align}
\end{widetext}In the absence of an electric field, the zeroth order
DM matrix element is given by the Fermi-Dirac distribution and the
band space identity matrix,
\begin{equation}
\rho_{\mathbf{k}ss'}^{(0)}=f(\epsilon_{\mathbf{k}s})\,\delta_{ss'}.
\end{equation}
The equivalence between these two conductivities, as well as for conductivities
at an arbitrary order $n$, is ensured by the existence of sum rules
\citep{Sipe1993,passos}, that are valid as long as the integration
over $\mathbf{k}$ is taken over the \textit{full} FBZ. Though it
is still possible to perform calculations using only a \textit{portion}
of the FBZ \textemdash{} e.g., graphene in the Dirac cone approximation
\citep{Cheng2014,Cheng2015,Cheng2015b,Mikhailov2016,Mikhailov2017}
\textemdash{} one must do so in the length gauge \citep{ventura},
making it the suitable choice for analytical calculations \citep{passos}.
In this work we present two analytical results, for the effects of
second harmonic generation and optical rectification, in the clean
limit: $\gamma\rightarrow0$. 
\par\end{flushleft}

\begin{flushleft}
The velocity gauge, on the other hand, is suitable for numerical approaches
that involve the entire FBZ \citep{passos}. It does not feature higher
order poles, and it avoids having to take derivatives of the density
matrix. Instead, for a response of order $n$, one has to compute
all $h$ coefficients, Eq.(\ref{eq:AITCH}), up to order $n+1$, $h_{\mathbf{k}ss'}^{\alpha_{1}...\alpha_{n+1}}$.
All numerical results in this paper were calculated in the velocity
gauge.
\par\end{flushleft}

\section{velocity gauge for Tight-Binding Hamiltonians\label{sec:III}}

A perturbative description of the response in the velocity gauge is
correct only if the unperturbed Hamiltonian is defined in the full
FBZ. For the purpose of this work, we consider it to be a tight-binding
model. This section is therefore dedicated to two points that concern
this type of Hamiltonian: we show that the calculation of $h$ coefficients
is made simple when one chooses a basis for which all Berry connections
are trivial; we also trace the source of variations in the calculations
of these coefficients, and to the Berry connections found in the literature,
to subtle changes in the definition of the position operator. This
difference is illustrated in the linear optical response of the plain
graphene monolayer.

\subsection{Covariant derivatives in the second Bloch basis}

A tight-binding model is a simplified description of electrons in
a lattice, where electronic motion is characterized by hoppings from
one orbital to its neighbouring ones ($t_{ij}\left(\mathbf{R}_{n},\mathbf{R}_{m}\right)$),
where $i,j$ index different orbitals of the same unit cell, which
may have distinct on-site energies ($\epsilon_{i}$). In real space,
\begin{align}
\mathcal{H}= & \underset{\mathbf{R}_{n},\,\mathbf{R}_{n},\,i,\,j}{\sum\sum}\left[\,t_{ij}(\mathbf{R}_{n},\mathbf{R}_{m})\bigl|\phi_{\mathbf{R}_{n}\,i}\bigr\rangle\bigl\langle\phi_{\mathbf{R}_{m}\,j}\bigr|+\text{h.c.}\right]\nonumber \\
 & +\sum_{\mathbf{R}_{n}}\sum_{i}\,\epsilon_{i}(\mathbf{R}_{n})\bigl|\phi_{\mathbf{R}_{n}\,i}\bigr\rangle\bigl\langle\phi_{\mathbf{R}_{n}\,i}\bigr|.
\end{align}
 A $\bigl|\phi_{\mathbf{R}_{n}\,i}\bigr\rangle$ represents a Wannier
orbital centered at the position $\mathbf{R}_{n}+\boldsymbol{\lambda}_{i}$,
with $\boldsymbol{\lambda}_{i}$ being the vector from that $i$-orbital
site to the unit cell origin.

As seen in the previous section, the eigenvalues of this Hamiltonian
are the bands, $\epsilon_{\mathbf{k}s}$, and the eigenfunctions are
the Bloch eigenstates, $\bigl|\psi_{\mathbf{k}s}\bigr\rangle$. There
is, however, a second basis of functions that also satisfies Bloch's
theorem, where each Bloch state is built out of a single type of Wannier
orbitals (same $i$),
\begin{align}
\bigl|\psi_{\mathbf{k}i}\bigr\rangle= & \sum_{\mathbf{R}_{n}}e^{i\mathbf{k}\cdot\left(\mathbf{R}_{n}+\boldsymbol{\lambda}_{i}\right)}\bigl|\phi_{\mathbf{R}_{n}\,i}\bigr\rangle.\label{eq:LOCAL BLOCH}
\end{align}
A very common approximation is to \emph{define} the position operator
as diagonal in the Wannier basis:
\begin{align}
\mathbf{r}\bigl|\phi_{\mathbf{R}_{n}\,i}\bigr\rangle= & \left(\mathbf{R}_{n}+\boldsymbol{\lambda}_{i}\right)\bigl|\phi_{\mathbf{R}_{n}\,i}\bigr\rangle.\label{eq:POSITION WANNIER}
\end{align}
Under this approximation, the periodic factor in the Bloch wavefunction
\[
\bigl|u_{\mathbf{k}i}\bigr\rangle=e^{-i\mathbf{k}\cdot\mathbf{r}}\bigl|\psi_{\mathbf{k}i}\bigr\rangle=\sum_{\mathbf{R}_{n}}\bigl|\phi_{\mathbf{R}_{n}\,i}\bigr\rangle,
\]
is $\mathbf{k}$ independent and the Berry connections in this second
basis is trivially zero,

\begin{equation}
\xi_{\mathbf{k}ij}^{\alpha}=i\left\langle u_{\mathbf{k}i}\right|\left.\nabla_{\mathbf{k}}u_{\mathbf{k}j}\right\rangle .\label{eq:LOCAL BERRY}
\end{equation}
This means that in the second Bloch basis, the covariant derivative
($\mathbf{D}_{\mathbf{k}}$) reduces to the regular derivative ($\nabla_{\mathbf{k}}$)
and that the matrix element of the derivative of an operator is simply
the derivative of matrix element of that operator \citep{ventura},
\begin{align}
\left\langle u_{\mathbf{k}i}\right|\left(\nabla_{\mathbf{k}}^{\alpha}\mathcal{O}_{\mathbf{k}}\right)\left|u_{\mathbf{k}j}\right\rangle = & \left[D_{\mathbf{k}}^{\alpha},\mathcal{O}_{\mathbf{k}}\right]_{ij},\nonumber \\
= & \nabla_{\mathbf{k}}^{\alpha}\left[\left\langle u_{\mathbf{k}i}\right|\mathcal{O}_{\mathbf{k}}\left|u_{\mathbf{k}j}\right\rangle \right]-i\left[\xi_{\mathbf{k}}^{\alpha},\,\mathcal{O}_{\mathbf{k}}\right]_{ij},\nonumber \\
= & \nabla_{\mathbf{k}}^{\alpha}\mathcal{O}_{\mathbf{k}ij}.
\end{align}
The calculation of $h$ coefficients is then fairly simple. Following
from Eq.(\ref{eq:AITCH}), and by use of the completeness relation
for the states in the second basis, we can see that
\begin{align}
h_{\mathbf{k}ss'}^{\alpha_{1}...\alpha_{p}}= & \sum_{i,j}\left\langle u_{\mathbf{k}s}\right|\left.u_{\mathbf{k}i}\right\rangle \left\langle u_{\mathbf{k}i}\right|\left(\nabla_{\mathbf{k}}^{\alpha_{1}}...\nabla_{\mathbf{k}}^{\alpha_{p}}H_{\mathbf{k}}\right)\bigl|u_{\mathbf{k}j}\bigr\rangle\left\langle u_{\mathbf{k}j}\right|\left.u_{\mathbf{k}s'}\right\rangle ,\nonumber \\
= & \sum_{i,j}c_{\mathbf{k}s,i}\left(\nabla_{\mathbf{k}}^{\alpha_{1}}...\nabla_{\mathbf{k}}^{\alpha_{p}}H_{\mathbf{k}ij}\right)c_{\mathbf{k}s',j}^{*}\,,\label{eq:AITCH_TB}
\end{align}
for, $c_{\mathbf{k}s,i}$, the solutions to the eigenvector problem
for that particular value of \textbf{k},
\begin{align}
\left|\psi_{\mathbf{k}s}\right\rangle = & \sum_{i}c_{\mathbf{k}s,i}\left|\psi_{\mathbf{k}i}\right\rangle .
\end{align}
The Berry connection, in particular, is
\begin{align}
\xi_{\mathbf{k}ss'}^{\alpha} & =i\left\langle u_{\mathbf{k}s}\right|\left.\nabla_{\mathbf{k}}u_{\mathbf{k}s'}\right\rangle ,\nonumber \\
 & =i\sum_{j,i}\left\langle u_{\mathbf{k}s}\right|\left.u_{\mathbf{k}j}\right\rangle \left\langle u_{\mathbf{k}j}\right|\left.u_{\mathbf{k}i}\right\rangle \left(\nabla_{\mathbf{k}}^{\alpha}\left\langle u_{\mathbf{k}i}\right|\left.u_{\mathbf{k}s'}\right\rangle \right),\\
 & =i\sum_{j}c_{\mathbf{k}s,j}\nabla_{\mathbf{k}}^{\alpha}c_{\mathbf{k}s',j}^{*},\label{eq:Berry_change_baisis}
\end{align}
since $\nabla_{\mathbf{k}}^{\alpha}\left(\left|u_{\mathbf{k}i}\right\rangle \left\langle u_{\mathbf{k}i}\right|\left.u_{\mathbf{k}s'}\right\rangle \right)=\left|u_{\mathbf{k}i}\right\rangle \nabla_{\mathbf{k}}^{\alpha}\left\langle u_{\mathbf{k}i}\right|\left.u_{\mathbf{k}s'}\right\rangle .$
Note that this procedure for computing the $h$ coefficients has a
profound impact on how the numerical calculations of the conductivity
are performed: by having $H_{\mathbf{k}ij}$ that are analytical,
one can easily compute their derivatives. All other operations, such
as solving the eigenvalue/eigenvector problem and calculating matrix
elements in the band basis, can be done numerically and much more
efficiently. 
\begin{figure}[t]
\centering{}\includegraphics[scale=0.45]{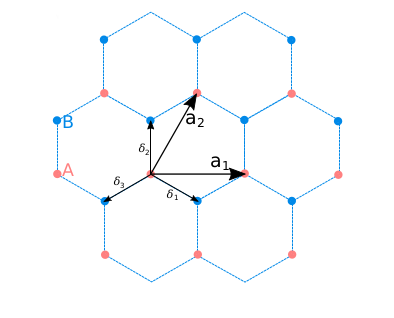}\caption{The honeycomb lattice of graphene with lattice parameter $\left|\boldsymbol{\delta}_{2}\right|=a_{0}$
and the armchair in the $\hat{y}$ direction. We have represented
both the lattice vectors, ($\boldsymbol{a}_{1}$, $\boldsymbol{a}_{2}$),
and the vectors connecting an A atom to its nearest neighbours, ($\boldsymbol{\delta}_{1}$,
$\boldsymbol{\delta}_{2}$, $\boldsymbol{\delta}_{3}$). In plain
graphene (PG), the atoms A and B are equivalent, in gapped graphene
(GG), these are not.\label{fig:honey}}
\end{figure}

\subsection{Choosing a representation for the tight-binding Hamiltonian}

There is a second issue concerning tight-binding Hamiltonians that,
though it does not pertain solely to the velocity gauge, is extremely
relevant in the calculation of nonlinear optical conductivities. For
simplicity, we present the following discussion in terms of the nearest
neighbour tight-binding model for the PG monolayer, since it will
be used in our description of the nonlinear optical responses. The
hopping parameter is set to $3$ eV, both in this section and throughout
the rest of the work \citep{passos,Hipolito2018}.

This Hamiltonian is usually written in two different ways,
\begin{align}
H_{\mathbf{k},(a/\delta)}= & \left[\begin{array}{cc}
0 & (-t)\ \phi_{(\delta/a)}(\mathbf{k})\\
(-t)\ \phi_{(\delta/a)}^{*}(\mathbf{k}) & 0
\end{array}\right].\label{eq:HAM_PG}
\end{align}
The first comes directly from the definition of the second Bloch basis
as that in Eq.(\ref{eq:LOCAL BLOCH}),
\begin{align}
\phi_{(\delta)}(\mathbf{k})= & \ e^{-i\mathbf{k}\cdot\boldsymbol{\delta}_{1}}+e^{-i\mathbf{k}\cdot\boldsymbol{\delta}_{2}}+e^{-i\mathbf{k}\cdot\boldsymbol{\delta}_{3}},\label{eq:F_DELTA}
\end{align}
and is expressed in terms of the vectors connecting an atom to its
nearest neighbours, Figure \ref{fig:honey}. The other way of writing
this Hamiltonian is associated to a second Bloch basis that has its
states phase shifted with respect to those of Eq.(\ref{eq:LOCAL BLOCH}),
\begin{align}
\bigl|\tilde{\psi}_{\mathbf{k}i}\bigr\rangle= & \sum_{\mathbf{R}_{n}}e^{i\mathbf{k}\cdot\mathbf{R}_{n}}\bigl|\phi_{\mathbf{R}_{n}\,i}\bigr\rangle,\label{eq:LOCAL BLOCH-1}
\end{align}
such that hoppings are written in terms of the lattice vectors $\mathbf{a}_{1}$
and $\mathbf{a}_{2}$, \citep{Cheng2014,Mikhailov2016,passos},
\begin{figure}[t]
\centering{}\includegraphics[scale=0.33]{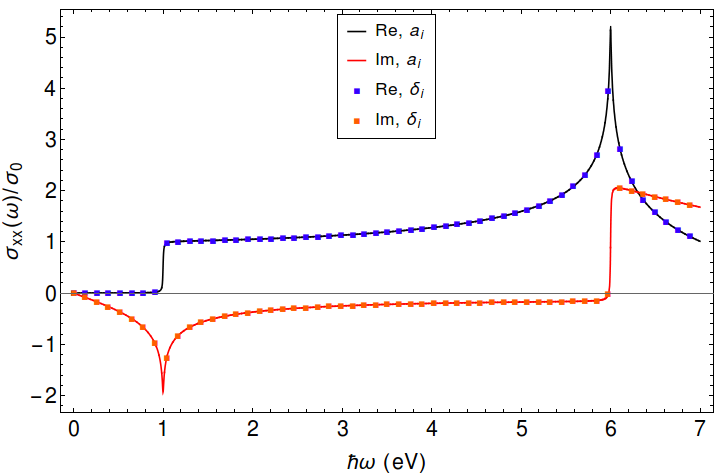}\caption{\textcolor{black}{The real and imaginary parts of the interband portion
of the linear conductivity, $\sigma_{xx}(\omega)$, for tight-binding
Hamiltonians written in the lattice vector ($a_{i}$) and nearest
neighbour ($\delta_{i}$) representations.\label{fig:2}} The relevant
parameters here are $\mu=0.5$ eV, $\gamma=0.005$ eV and $T=1$ K.
The conductivity is normalized with respect to $\sigma_{0}=e^{2}/4\hbar$.}
\end{figure}

\begin{align}
\phi_{(a)}(\mathbf{k})= & \ 1+e^{i\mathbf{k}\cdot(\mathbf{a}_{2}-\mathbf{a}_{1})}+e^{i\mathbf{k}\cdot\mathbf{a}_{2}}.\label{eq:F_A}
\end{align}
Both representations have the same eigenvalues since the $\phi$ functions
are related by a phase factor, 
\begin{equation}
\phi_{(a)}(\mathbf{k})=e^{i\mathbf{k}\cdot\boldsymbol{\delta}_{2}}\,\phi_{(\delta)}(\mathbf{k}).
\end{equation}
\textcolor{black}{There is, however, a very important subtlety. If
we use Eq.(\ref{eq:AITCH_TB}) to define the $h$ coefficients, or
Eq.(\ref{eq:Berry_change_baisis}) to compute the Berry connection
we obtain different results, both found in the literature, in the
two representations: Eqs.(\ref{eq:F_DELTA}) and (\ref{eq:F_A}).}

It would appear that the condition for a trivial Berry connection
in the entire FBZ, Eq.(\ref{eq:LOCAL BERRY}), that follows from the
definition of the position operator of Eq.(\ref{eq:POSITION WANNIER})
is satisfied in the $\psi_{\mathbf{k}i}$ basis but it is \textit{not}
satisfied in the $\tilde{\psi}_{\mathbf{k}i}$ basis, since 
\begin{equation}
\bigl|\tilde{u}_{\mathbf{k}i}\bigr\rangle=e^{-i\mathbf{k}\cdot\mathbf{r}}\bigl|\tilde{\psi}_{\mathbf{k}i}\bigr\rangle=\sum_{\mathbf{R}_{n}}e^{-i\mathbf{k}\cdot\boldsymbol{\delta}_{i}}\bigl|\phi_{\mathbf{R}_{n}\,i}\bigr\rangle\label{eq:u_tilde}
\end{equation}
Still one needs to point out Eqs.(\ref{eq:AITCH_TB}) and (\ref{eq:Berry_change_baisis})
are still valid for the $\bigl|\tilde{\psi}_{\mathbf{k}i}\bigr\rangle$
basis of Eq.(\ref{eq:LOCAL BLOCH-1}), \emph{provided we define $\mathbf{r}$
differently, }effectively neglecting the distances inside the unit
cell, $\left|\boldsymbol{\lambda}_{i}\right|$,\emph{
\begin{equation}
\mathbf{r}\bigl|\phi_{\mathbf{R}_{n}\,i}\bigr\rangle=\mathbf{R}_{n}\bigl|\phi_{\mathbf{R}_{n}\,i}\bigr\rangle.\label{eq:POSITIO-WANNIER-2}
\end{equation}
}

These two different representations of the position operator, correspond
naturally to different approximations to the perturbation term, and
can lead to different results. To illustrate these distinctions it
is worthwhile to compare the responses that follow from either representation,
under the consideration that $h$ coefficients can be computed following
Eq.(\ref{eq:AITCH_TB}), for both the zig-zag and armchair directions.

First, we consider the linear response along the zig-zag direction,
$\sigma_{xx}(\omega)$, represented in Figure \ref{fig:2}. 
\begin{figure}[t]
\centering{}\includegraphics[scale=0.32]{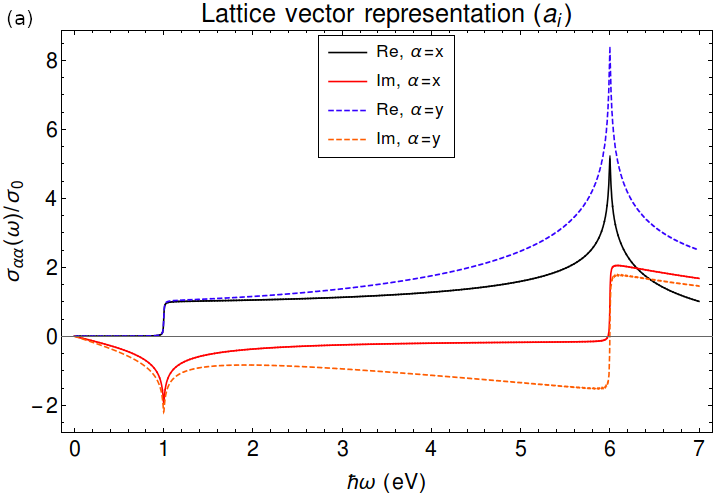}\smallskip{}
\includegraphics[scale=0.32]{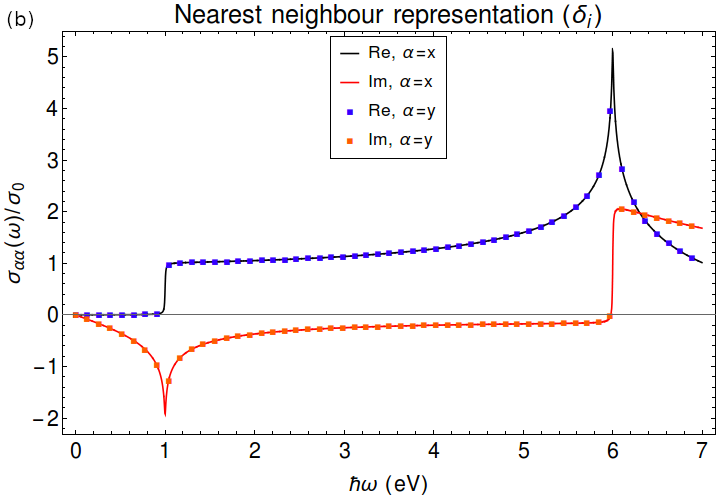}\caption{The real and imaginary parts of interband portion of the linear conductivities,
$\sigma_{xx}(\omega)$ and $\sigma_{yy}(\omega)$, following from
Eq.(\ref{eq:F_A})/(\ref{eq:F_DELTA}), the lattice vector representation
(a) and the nearest neighbour representation (b). The chemical potential
is again fixed to $0.5$ eV.\label{fig:3}}
\end{figure}
 In this case, the responses that follow from the two representations
are exactly the same, as both $\xi_{\mathbf{k}ij}^{x}$ and $\tilde{\xi}_{\mathbf{k}ij}^{x}$
are zero. This is due to the fact that $\boldsymbol{\delta}_{2}$
points in the $\hat{y}$ or armchair direction and as such, does not
bear an influence in the response along the zig-zag direction. For
the armchair direction, however, it is clear that the results in the
two representations are different, Figure \ref{fig:3}. More importantly,
we can see that in lattice vector representation, the responses along
zig-zag, $\sigma_{xx}(\omega)$, and armchair directions, $\sigma_{yy}(\omega)$,
are different from one another, Figure \ref{fig:3}(a). The use of
the approximation described in Eq.(\ref{eq:POSITIO-WANNIER-2}) fails
to properly translate the symmetry properties of the PG monolayer,
particularly at high frequencies \citep{Boyd:2008}. In the nearest
neighbours representation, that property is indeed fulfilled. It was
this latter representation that we used for the remaining numerical
results in this work.

\section{Results\label{sec:IV}}

In this section, we present the results for the second order response,
i.e. the second harmonic generation and optical rectification conductivities,
of the gapped graphene monolayer and for the third order response,
i.e. the third harmonic generation and Kerr effect conductivities,
of the plain and gapped monolayers to a monochromatic electric field.
We considered different values for the gap and chemical potential,
$\Delta$ and $\mu$, as well as different values for the scattering
rate, $\gamma$, but not different values for the temperature, $T$,
as its effect is similar to that of $\gamma$, which is to broaden
the features. $T$ is thus set, throughout this work, to $1$K. Since
all nonlinear optical conductivities are monotonically decreasing
\textemdash{} the exception being the regions around processes at
the gap (or twice the chemical potential) and around the van Hove
singularities \textemdash{} these were represented in the two frequency
regions separately, so as to make the features more visible. It must
be said of these high frequency results that they should be taken
only as an indication of what the response should look like \textemdash{}
they were calculated in the independent particle approximation and,
as such, do not consider the effect of excitons \textcolor{black}{\citep{Chae2011,Mak2014}}.
Finally, we emphasize that the following conductivities satisfy the
property of intrinsic permutation symmetry \citep{Boyd:2008}.

\subsection{SECOND ORDER RESPONSE OF THE GAPPED GRAPHENE MONOLAYER}

A gap is introduced in the plain graphene Hamiltonian, Eq.(\ref{eq:HAM_PG}),
by adding to it a term that breaks the equivalence between the A and
B atoms, $\text{diag}(\Delta/2,-\Delta/2)$, and thus the centrosymmetricity
of the PG monolayer. The study of the remaining symmetries in the
point group then tells us that there is only one relevant component
for this conductivity tensor: $\sigma_{yyy}$, with $y$ being the
armchair direction \citep{Hipolito2016}. In addition, the relation
between this component and the remaining nontrivial components reads
as,
\begin{equation}
\sigma_{xxy}=\sigma_{xyx}=\sigma_{yxx}=-\sigma_{yyy}.\label{eq:TENSOR}
\end{equation}
The following results have been normalized with respect to $\sigma_{2}=e^{3}a_{0}/4t\hbar=2.87\times10^{-15}\text{ S\ensuremath{\cdot}m/V}$
\citep{Hipolito2016}.

\subsubsection{Second Harmonic Generation (SHG)}

\begin{figure}[t]
\centering{}\includegraphics[scale=0.3]{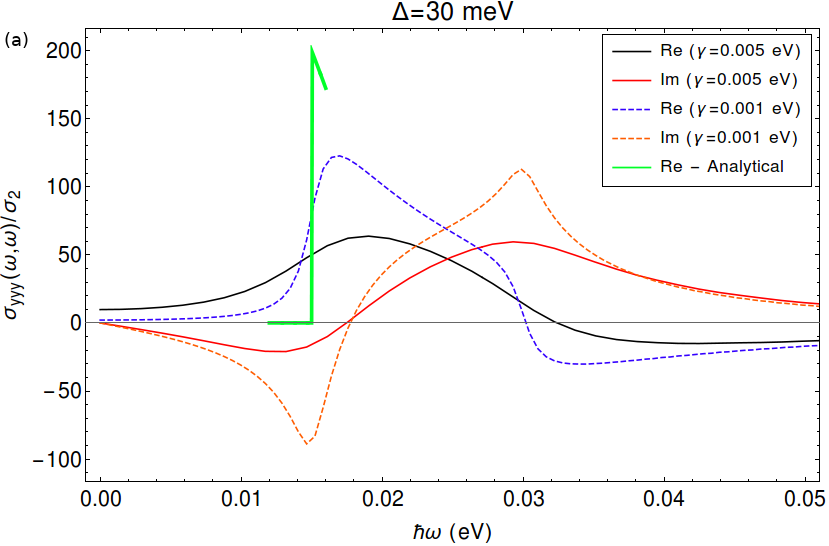}$\ \,$\smallskip{}
\includegraphics[scale=0.3]{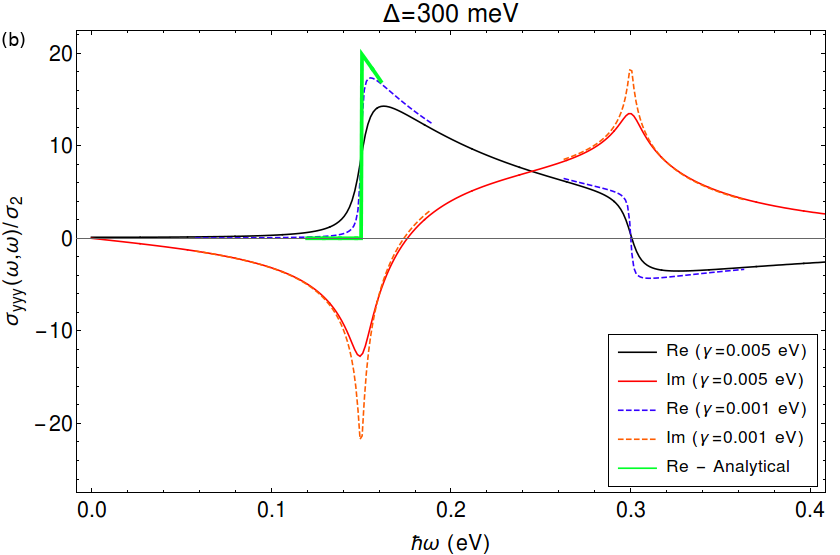}\caption{The real and imaginary parts of the second harmonic generation, $\sigma_{yyy}(\omega,\omega)$,
close to the one photon and two photon processes at the gap: $30$
meV in the top plot, (a), and $\Delta=300$ meV in the bottom plot,
(b) for different values of the scattering rate. The green curve represents
the $\gamma=0$ analytical result. The parameters not listed in the
plots are the chemical potential and the temperature, which will be
fixed to $\mu=0$ eV, $T=1$K. These values are used in the remaining
figures of this section. \label{fig:4}}
\end{figure}

We begin with the study of the one photon ($\hbar\omega\sim\Delta$)
and two photon ($2\hbar\omega\sim\Delta$) processes at the gap for
values of $\Delta=30,\,300$ meV \citep{Hipolito2018} and for two
different values of the scattering rate, $\gamma=0.005,\,0.001$ eV,
Figure \ref{fig:4}.
 It is clear from these results that the shape of the features is
highly dependent on the interplay between the gap and scattering parameters,
$\gamma$ and $\Delta$. For the larger scattering rate and smaller
gap, we can see an overlap of the two and one photon peaks, Figure
\ref{fig:4}(a), which is markedly different from what happens for
the larger gap, Figure \ref{fig:4}(b), where the two peaks are clearly
distinct. For smaller values of the scattering rate, represented by
dashed curves, there is a sharpening of the features \textemdash{}
now narrower and taller \textemdash{} and the results for the two
gaps are similar. To study the zero scattering limit, $\gamma=0$,
we turn to the analytical results \textemdash{} represented by the
thicker green curve \textemdash{} that are obtained in the length
gauge, Eq.(\ref{eq:SIGMA_E}). It can be shown that the real part
of the two photon process in the second harmonic generation can be
expressed in terms of the shift current coefficient that has been
previously derived in \citep{Aversa1995,Sipe2000},
\begin{align}
\frac{\text{Re}\left[\sigma_{yyy}(\omega,\omega)\right]}{\sigma_{2}}= & -\frac{8it}{\pi}\int d^{2}\mathbf{k}\,\xi_{vc}^{y}\bigl(\xi_{cv}^{y}\bigr)_{;y}\ \delta(2\hbar\omega-\Delta\epsilon_{cv}).\label{eq:SHG_ANALYTICAL}
\end{align}
using the standard notation for the generalized derivative, $\bigl(\xi_{ss'}^{\alpha_{1}}\bigr)_{;\alpha_{2}}=\nabla_{\mathbf{k}}^{\alpha_{1}}(\xi_{ss'}^{\alpha_{2}})-i(\xi_{ss}^{\alpha_{1}}-\xi_{s's'}^{\alpha_{1}})\xi_{ss'}^{\alpha_{2}}$
\citep{Aversa1995}. By having a delta function in the integrand,
one can see that the relevant contributions to the study of the two
photon processes at the gap will come from the two regions of the
FBZ around the band minimum, $\mathbf{K},\,\mathbf{K}'=\pm4\pi/3\sqrt{3}a_{0}\,\hat{x}$,
which motivates a momentum expansion of the band around those points.
Furthermore, since the delta function fixes $\Delta\epsilon_{cv}$
directly to twice the photon energy it is the suitable variable of
integration,
\begin{align}
\Delta\epsilon_{cv}^{2}= & \Delta^{2}+4t^{2}\left|\phi_{\delta}(\mathbf{k})\right|^{2}.\label{eq:SQ_DIS_REL}
\end{align}
Now, by expanding the hopping function, $\phi_{\delta}$, for small
momenta around one of the band minima, 
\begin{align}
\left|\phi_{\delta}(\mathbf{k}=\mathbf{K}+\mathbf{q})\right|= & \frac{3\left|\mathbf{q}\right|}{2}-\frac{3\left|\mathbf{q}\right|^{2}}{8}\text{cos}(3\theta)+\mathcal{O}(\left|\mathbf{q}\right|^{3}),\label{eq:EXPANSION}
\end{align}
where $\left|\mathbf{q}\right|$ and $\theta$ are the radial and
polar coordiantes associated with $\mathbf{q}$, and by rewriting
Eq.(\ref{eq:SQ_DIS_REL}) with the help of Eq.(\ref{eq:EXPANSION}),
we obtain,
\begin{align}
\frac{1}{t}\sqrt{\Delta\epsilon_{cv}^{2}-\Delta^{2}}= & \frac{3\left|\mathbf{q}\right|}{2}-\frac{3\left|\mathbf{q}\right|^{2}}{8}\text{cos}(3\theta)+\mathcal{O}(\left|\mathbf{q}\right|^{3}).\label{eq:EQUAL}
\end{align}
We have effectively related one of our integration variables, $\left|\mathbf{q}\right|$,
with the small parameter $\delta(\Delta\epsilon_{cv})=\sqrt{\Delta\epsilon_{cv}^{2}-\Delta^{2}}/t$.
It is now possible to invert this series, so as to obtain $\left|\mathbf{q}\right|$
in terms of $\delta$,
\begin{figure}[t]
\centering{}\smallskip{}
\smallskip{}
\includegraphics[scale=0.265]{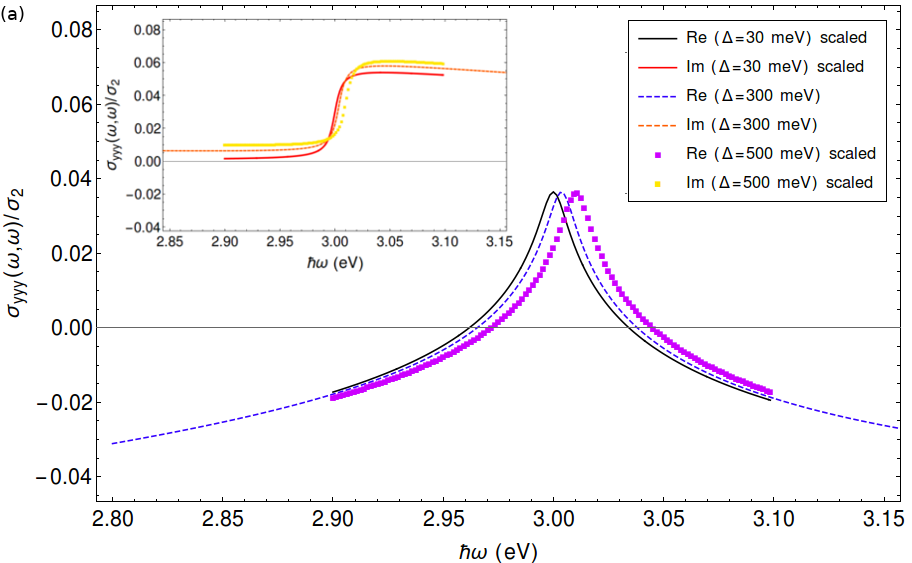}\smallskip{}
\includegraphics[scale=0.26]{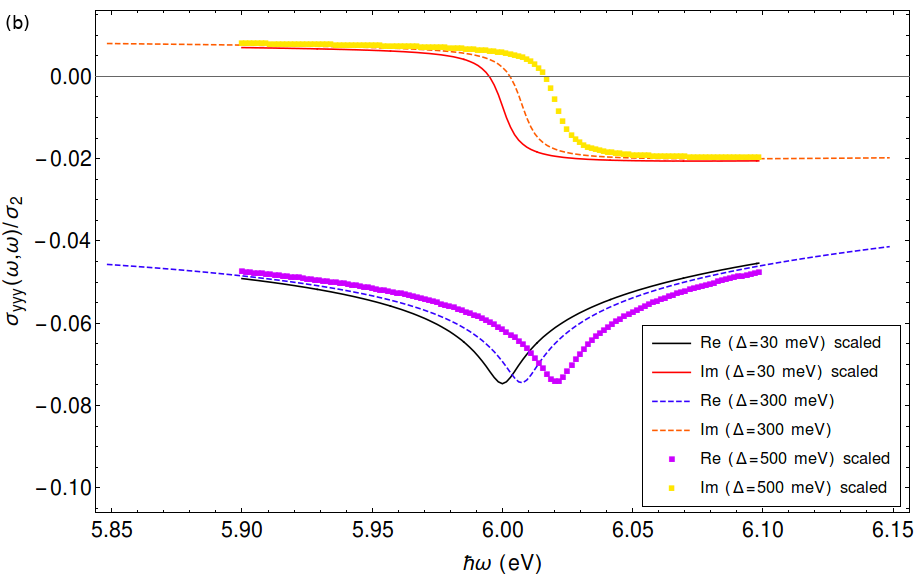}\caption{The real and imaginary parts of the SHG in GG, $\sigma_{yyy}(\omega,\omega)$,
for frequencies around the two ($\hbar\omega\sim t$) (a) and one
($\hbar\omega\sim2t$) (b) photon processes at the van Hove singularity.
The imaginary parts for the two photon processes are represented on
the inset. Curves labelled as scaled have been divided by a factor
of $\Delta/300$ meV. The scattering parameter for these plots is
$\gamma=0.005$ eV. \label{fig:5}}
\end{figure}
\begin{align}
\left|\mathbf{q}\right| & =\frac{\delta}{3}+\frac{\delta^{2}}{36}\text{cos}(3\theta)+\mathcal{O}(\delta^{3}).
\end{align}
Performing this change of variable in the integral, $\left|\mathbf{q}\right|\rightarrow\delta$,
enables us to compute the integration in Eq.(\ref{eq:SHG_ANALYTICAL})
analytically. The result is an expansion in powers of $(2\hbar\omega)^{2}-\Delta^{2}$,
which in the lowest orders reads,

\begin{align}
\frac{\text{Re}\left[\sigma_{yyy}(\omega,\omega)\right]}{\sigma_{2}}=\  & \Theta(2\hbar\omega-\Delta)\Bigl[\frac{2t}{\Delta}+\Bigl(\frac{t}{9\Delta}-\frac{2t^{3}}{\Delta^{3}}\Bigr)\nonumber \\
 & \times\Bigl(\Bigl(\frac{2\hbar\omega}{t}\Bigr)^{2}-\Bigl(\frac{\Delta}{t}\Bigr)^{2}\Bigr)+(...)\Bigr].\label{eq:TPA_ANA}
\end{align}
This represented in Figure \ref{fig:4}, alongside the numerical results
of the velocity gauge. We must note that, had we carried only linear
terms in $\left|\mathbf{q}\right|$ in the expansion of the hopping
function, the $\text{Re}\left[\sigma_{yyy}(\omega,\omega)\right]$
would be exactly zero, for the same reason it vanishes in the monolayer
of plain graphene: in that case, the Berry connections, $\xi_{\mathbf{q}ss'}^{\alpha}$,
are odd under $\mathbf{q}\rightarrow-\mathbf{q}$, and the integral
vanishes necessarily. To obtain a nontrivial second order response
in GG one has to consider the trigonal warping terms in the expansion
of Eq.(\ref{eq:EXPANSION}). The high frequency results, i.e. those
for the two ($\hbar\omega\sim t$) and one ($\hbar\omega\sim2t$)
photon processes at the van Hove singularities, are represented in
Figure \ref{fig:5}. We can see that the features \textemdash{} for
different values of $\Delta$ \textemdash{} are centered around slightly
different different energies, as 
\begin{align}
\Delta\epsilon_{\textsc{vHs}}^{2}= & \Delta^{2}+4t^{2}\left|\phi_{\delta}(\mathbf{M})\right|^{2}.\label{eq:VHS}
\end{align}
Note also that the absolute value of these conductivities scales with
$\Delta$ \textemdash{} the opposite behavior to what we found for
the response at the gap. Another, quite surprising, point concerns
the features for the real and imaginary parts of these conductivities
as they are switched with respect to the real and imaginary parts
of the conductivities at the gap, Figure \ref{fig:4}. It is now the
real part that has the shape of a logarithmic-like divergence while
the step-like behavior is present in the imaginary part. 

\subsubsection{Optical Rectification}

\begin{figure}[t]
\centering{}\includegraphics[scale=1.27]{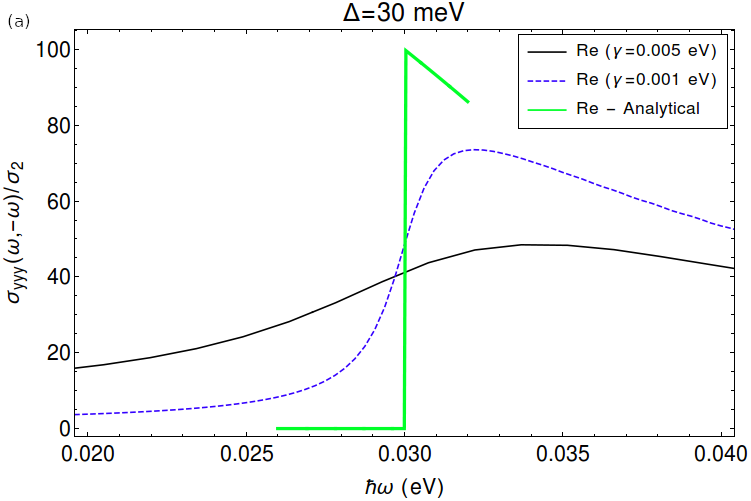}\smallskip{}
\includegraphics[scale=1.13]{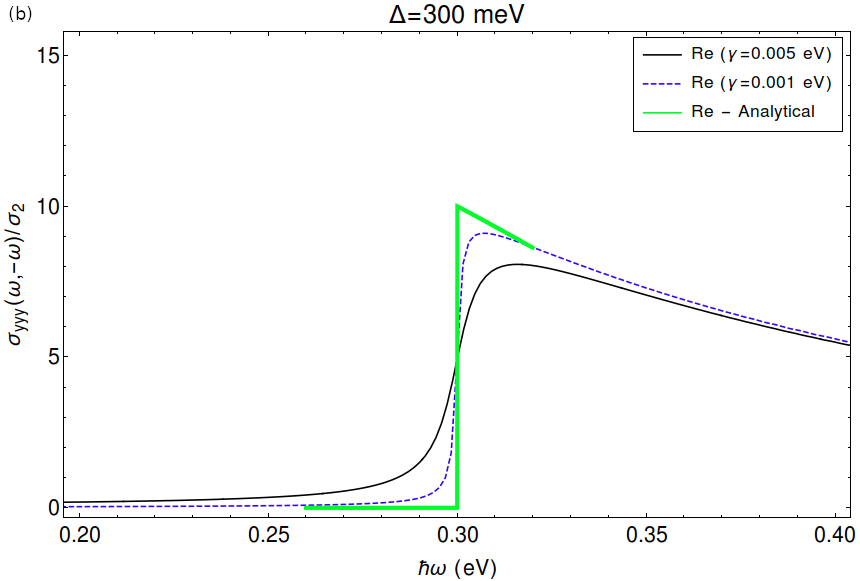}\caption{The optical rectification conductivity, $\sigma_{yyy}(\omega,-\omega)$,
in GG for frequencies close to the gap: $\Delta=30$ meV in the top
plot (a) and $\Delta=300$ meV in the bottom plot (b) for different
values of the scattering rate. The green curve represents the $\gamma=0$
analytical result of Eq.(\ref{eq:PHOTO_ANA}). \label{fig:6}}
\end{figure}

\begin{figure}[t]
\centering{}\includegraphics[scale=0.32]{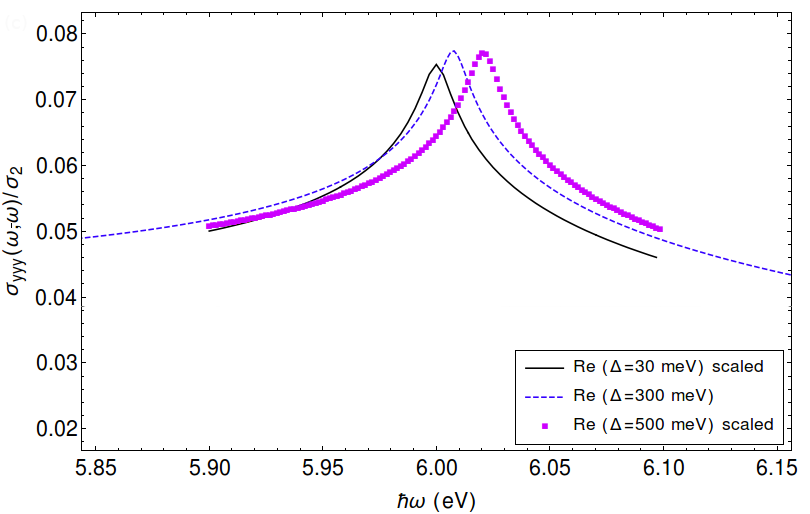}\caption{The optical rectification conductivity, $\sigma_{yyy}(\omega,-\omega)$,
in GG for frequencies around the one photon process at the van Hove
singularity. Curves labelled as scaled have been divided by a factor
of $\Delta/300$ meV. The scattering parameter in this plot is $\gamma=0.005$
eV. \label{fig:7}}
\end{figure}

The other second order process that can be observed in the response
to an external monochromatic field is the generation of a DC current,
described by the optical recitification conductivity: $\sigma_{yyy}(\omega,-\omega)$,
Figures \ref{fig:6} and \ref{fig:7}. 

From the inspection of the response at photon energies close to the
value of the gap, $\hbar\omega\sim\Delta$, Figure \ref{fig:6}, we
can see that this tensor component is always finite (even in the zero
scattering limit), meaning that there is indeed the absence of the
injection current, as prescribed by the symmetry properties of the
GG monolayer, Eq.(\ref{eq:TENSOR}). The remaining portion of this
response is associated to the shift current and has a feature which
is similar to that of the second harmonic generation at the gap, Figure
\ref{fig:4}(b). In the zero scattering limit, we have \citep{Sipe2000},
\begin{align}
\frac{\sigma_{yyy}(\omega,-\omega)}{\sigma_{2}}= & -\frac{4it}{\pi}\int d^{2}\mathbf{k}\,\xi_{vc}^{y}\bigl(\xi_{cv}^{y}\bigr)_{;y}\ \delta(\hbar\omega-\Delta\epsilon_{cv}).
\end{align}
By comparison with the two photon resonance in the second harmonic
generation, Eq.(\ref{eq:TPA_ANA}), we can see the two effects are
essentially described by the same function with just different arguments
\textemdash{} $2\omega$ in the case of the SHG \textemdash{} and
an extra factor of two,\footnote{We have compared this with the analytical result of ref.\citep{Hipolito2016}
and found a minus sign discrepancy, which \textemdash{} according
to our calculations \textemdash{} follows from excluding the derivative
portion of the generalized derivative.} 
\begin{align}
\frac{\text{Re}\left[\sigma_{yyy}(\omega,-\omega)\right]}{\sigma_{2}}=\  & \Theta(\hbar\omega-\Delta)\Bigl[\frac{t}{\Delta}+\Bigl(\frac{t}{18\Delta}-\frac{t^{3}}{\Delta^{3}}\Bigr)\nonumber \\
 & \times\Bigl(\Bigl(\frac{\hbar\omega}{t}\Bigr)^{2}-\Bigl(\frac{\Delta}{t}\Bigr)^{2}\Bigr)+(...)\Bigr].\label{eq:PHOTO_ANA}
\end{align}
The similarity between the optical rectification conductivity and
the second harmonic generation is also present at higher frequencies,
$\hbar\omega\sim2t$, Figure \ref{fig:7}. Apart from a sign switch
and the absence of the imaginary part \textemdash{} the symmetrized
optical rectification conductivity is necessarily real \textemdash{}
this result is very similar that of Figure \ref{fig:5}(b). 

\subsection{THIRD ORDER RESPONSE OF THE GG AND PG MONOLAYERS}

The third order response is finite even in the presence of inversion
symmetry and as such, we present results for both the gapped graphene
and the plain graphene monolayers for the nonlinear processes of third
harmonic generation and optical Kerr effect. Though associated to
different point group symmetries, the components of their third order
conductivities satisfy the same relations,
\begin{equation}
\begin{array}{c}
\sigma_{yyyy}=\sigma_{xxxx}=\sigma_{xxyy}+\sigma_{xyxy}+\sigma_{xyyx},\\
\sigma_{xxyy}=\sigma_{yyxx},\ \sigma_{xyxy}=\sigma_{yxyx},\ \sigma_{xyyx}=\sigma_{yxxy}.
\end{array}
\end{equation}
As we are also imposing intrinsic permutation symmetry, there is only
one relevant component in third harmonic generation (THG), with all
other components of the tensor trivially expressed in terms of it,
\begin{equation}
\begin{array}{c}
\sigma_{xxyy}(\omega,\omega,\omega)=\sigma_{xyxy}(\omega,\omega,\omega)=\sigma_{xyyx}(\omega,\omega,\omega),\\
\sigma_{xxyy}(\omega,\omega,\omega)=\frac{1}{3}\sigma_{xxxx}(\omega,\omega,\omega)=\frac{1}{3}\sigma_{yyyy}(\omega,\omega,\omega).
\end{array}\label{eq:THG_TENSOR_COM}
\end{equation}
We will thus present only the $\sigma_{yyyy}$ in our study of the
THG. For the optical Kerr effect, we consider both the $\sigma_{yyyy}$
and $\sigma_{yxxy}$ components. The following results have been normalized
by $\sigma_{3}=e^{4}a_{0}^{2}/8\hbar t^{2}=6.84\times10^{-26}\text{ S\ensuremath{\cdot}m}^{2}\text{/V}^{2}$.

\subsubsection{Third Harmonic Generation (THG)}

\begin{figure}[t]
\centering{}\includegraphics[scale=0.29]{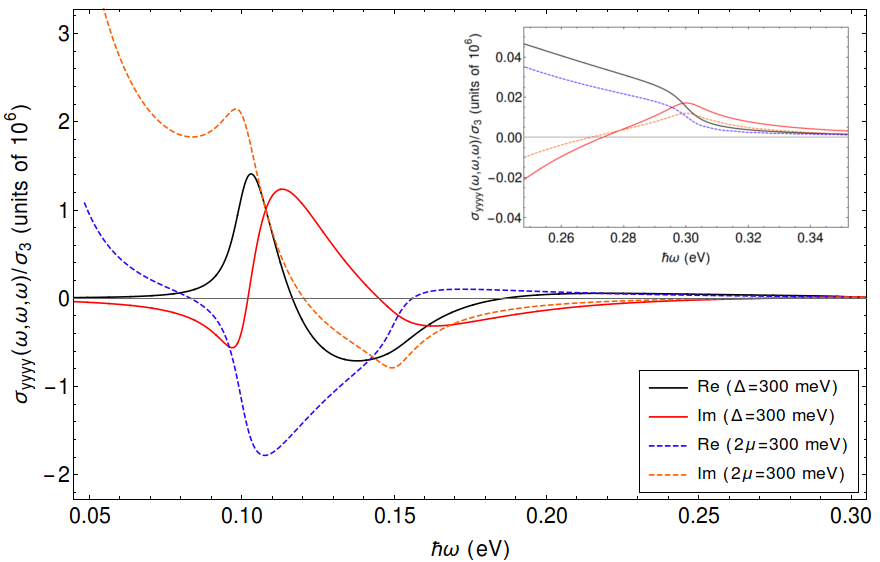}\caption{The real and imaginary parts of the third harmonic generation, $\sigma_{yyyy}(\omega,\omega,\omega)$,
in the GG, $\Delta=300$ meV, and PG, $2\mu=300$ meV, monolayers
for frequencies that cover the different (one, two and three) photon
processes at the gap / twice the value of the chemical potential.
Note that the vertical scale is in units of $10^{6}$. The inset represents
a zoom-in in the region of the one photon process, $\hbar\omega\sim\Delta,\,2\mu$.
\label{fig:8}}
\end{figure}
\begin{figure}[t]
\begin{centering}
\includegraphics[scale=0.26]{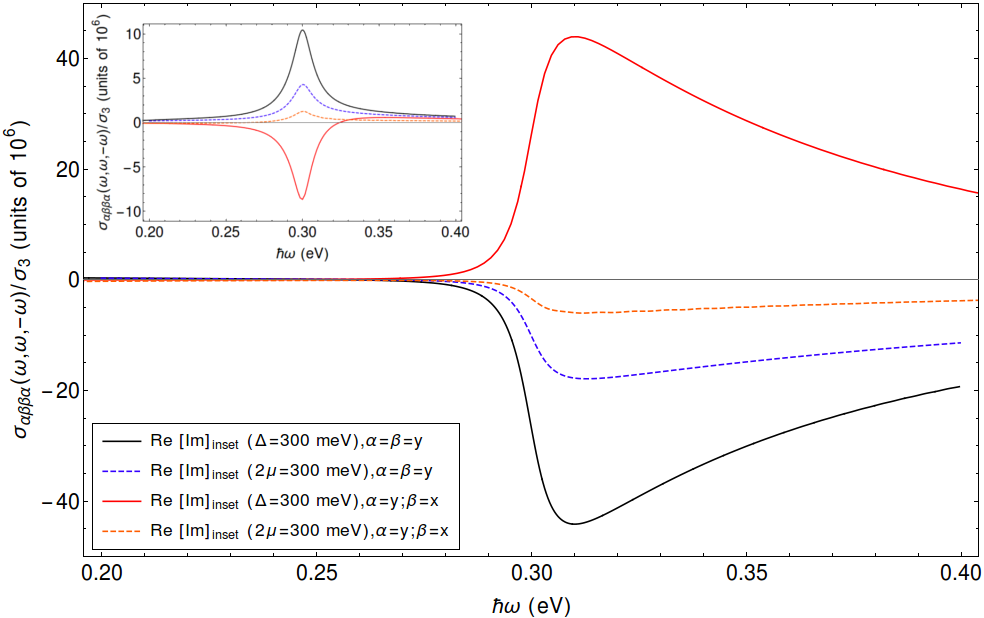}
\par\end{centering}
\caption{The real and imaginary parts (the latter is represented in the inset)
of the optical Kerr effect for the components, $\sigma_{yyyy}(\omega,\omega,-\omega)$
and $\sigma_{yxxy}(\omega,\omega,-\omega)$, in the GG, $\Delta=300$
meV, and PG, $2\mu=300$ meV, monolayers for frequencies around the
one photon process at the gap / twice the value of the chemical potential.
Note that the vertical scale is in units of $10^{6}$.\label{fig:fig9}}
\end{figure}
\begin{figure}[t]
\centering{}\includegraphics[scale=0.275]{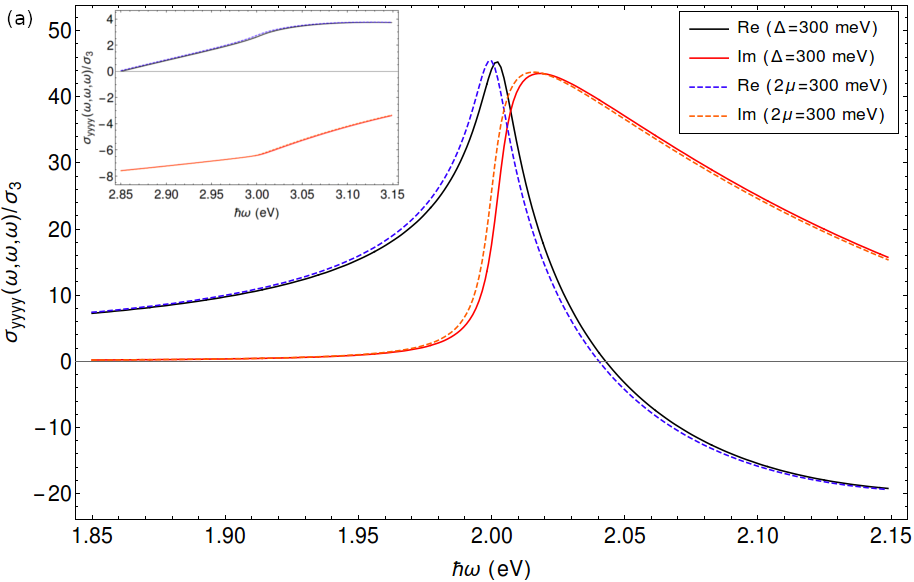}\smallskip{}
\includegraphics[scale=0.259]{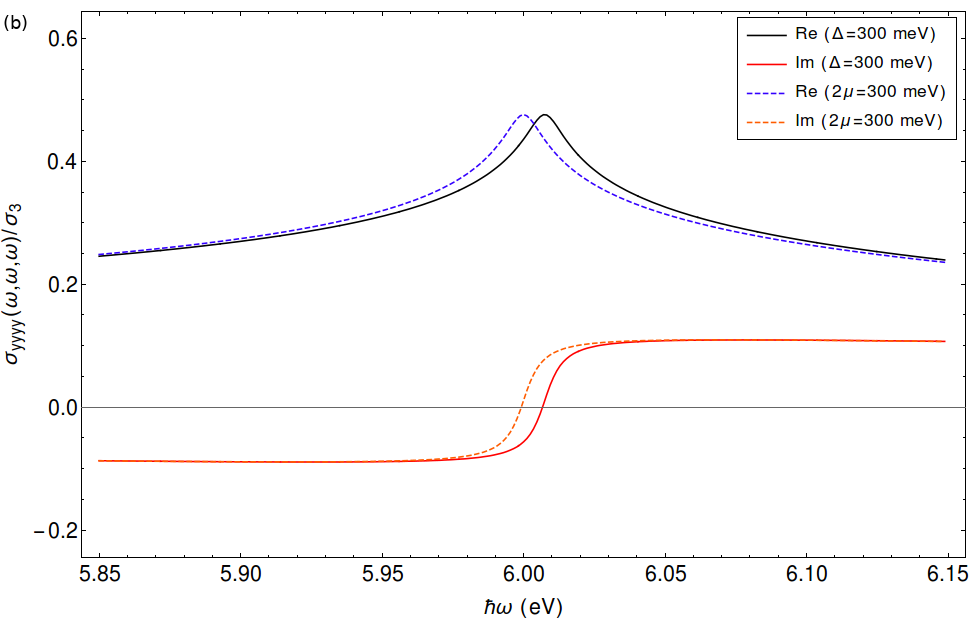}\caption{The real and imaginary parts of the third harmonic generation, $\sigma_{yyyy}(\omega,\omega,\omega)$,
for frequencies around the three photon ($3\hbar\omega\sim2t$), (a),
two photon ($\hbar\omega\sim t$), inset of (a), and one photon ($\hbar\omega\sim2t$),
(b), processes at the van Hove singularity in the GG, $\Delta=300$
meV, and PG, $2\mu=300$ meV, monolayers.\label{fig:9}}
\end{figure}

One of the points covered in a previous subsection (SHG) concerned
the interplay between $\Delta$ and $\gamma$, and how this affected
the features one sees in the conductivities. If one were to do this
analysis in the THG, one would again conclude that for larger scattering
rates one sees a broadening, possibly even a merger, of the main features
in the conductivity. The scattering rate is therefore fixed to $\gamma=0.005$
eV. We will, instead, focus on the THG of the gapped and plain graphene
monolayers, in the case where the value of the gap in the GG is equal
to the energy value of the region of states that are Pauli-blocked,
$2\mu$, of the PG, and that this is equal to $300$ meV, Figures
\ref{fig:8} and \ref{fig:9}. 

We begin by studying the response of the several different photon
processes, $n\hbar\omega=\Delta,\,2\mu$ for $n=1,2,3$, Figure \ref{fig:8}.
It is clear that the conductivities for gapped and plain graphene
are very much different: for the three photon resonance, there are
prominent features in both sets of curves but the sign appears to
be switched with respect to one another; for the two photon resonance,
there are no clear features in the GG monolayer, whereas in the PG,
one finds a shoulder and a local minimum in the real and imaginary
parts, respectively. An exception to this, however, are the features
for the one photon process, inset of Figure \ref{fig:8}. The differences
between the low frequency limit of the gapped and plain graphene monolayer
can be easily ascribed to the intraband terms of the response, dominant
in this frequency range, that are completely absent from the response
of the GG \textemdash{} a cold semiconductor \textemdash{} but present
in the response of the doped PG monolayer. 

For higher frequencies, associated with the different processes around
the van Hove singularities, Figure \ref{fig:9}, we can see that the
conductivities of the PG and GG monolayers are rather similar. For
those energies, the band structures are rather similar (as $\Delta\ll t$)
and the chemical potential that is set in the PG is completely irrelevant.
The only difference in the two curves comes from the different energy
values for the van Hove singularity, Eq.(\ref{eq:VHS}). 
 
\begin{figure*}[t]
\centering{}\includegraphics[scale=0.265]{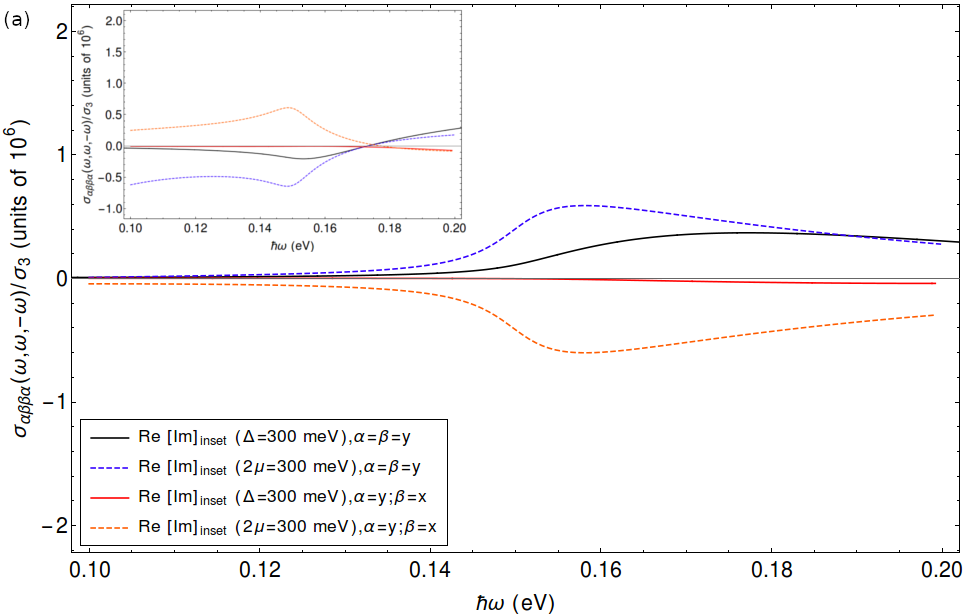}$\ \ \ $\includegraphics[scale=0.274]{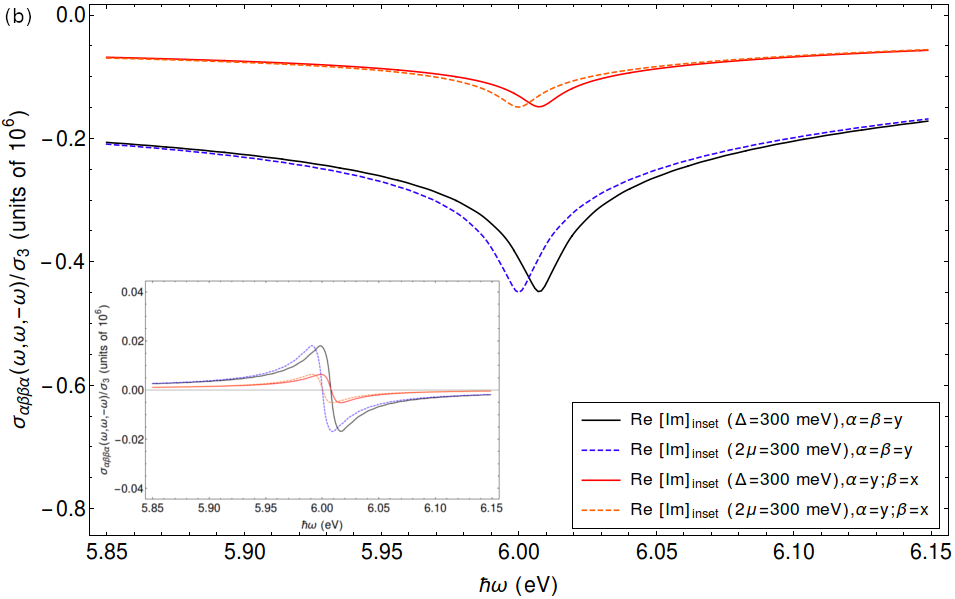}\caption{The real and imaginary (the latter represented in the insets) parts
of the Kerr effect, $\sigma_{yyyy}(\omega,\omega,-\omega)$, in the
GG, $\Delta=300$ meV, and PG, $2\mu=300$ meV, monolayers for frequencies
around $\Delta=\,2\mu$, (a), and for frequencies around the van Hove
singularity, (b). Note that the vertical scale in both figures is
in units of $10^{6}$. \label{fig:10} }
\end{figure*}
\begin{figure*}[t]
\centering{}\includegraphics[scale=0.28]{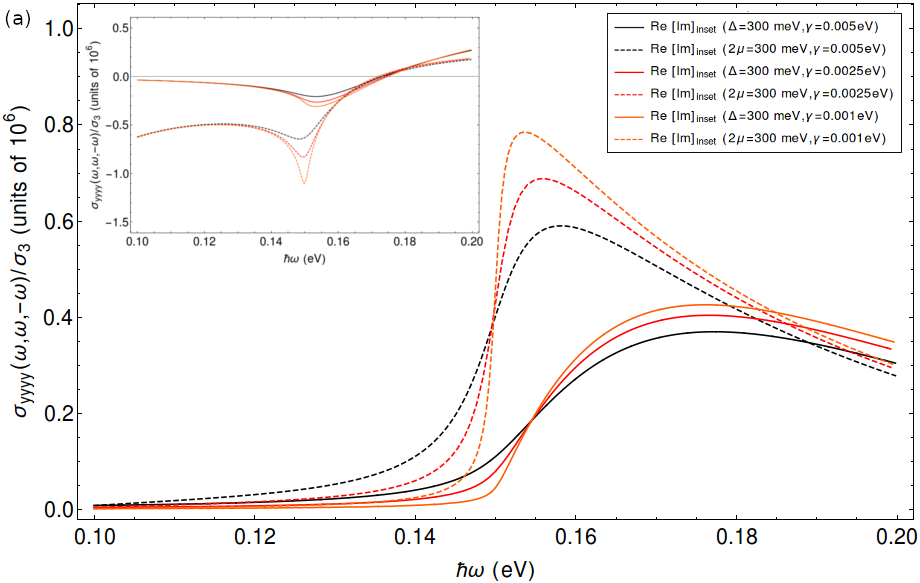}$\ \ \ $\includegraphics[scale=0.28]{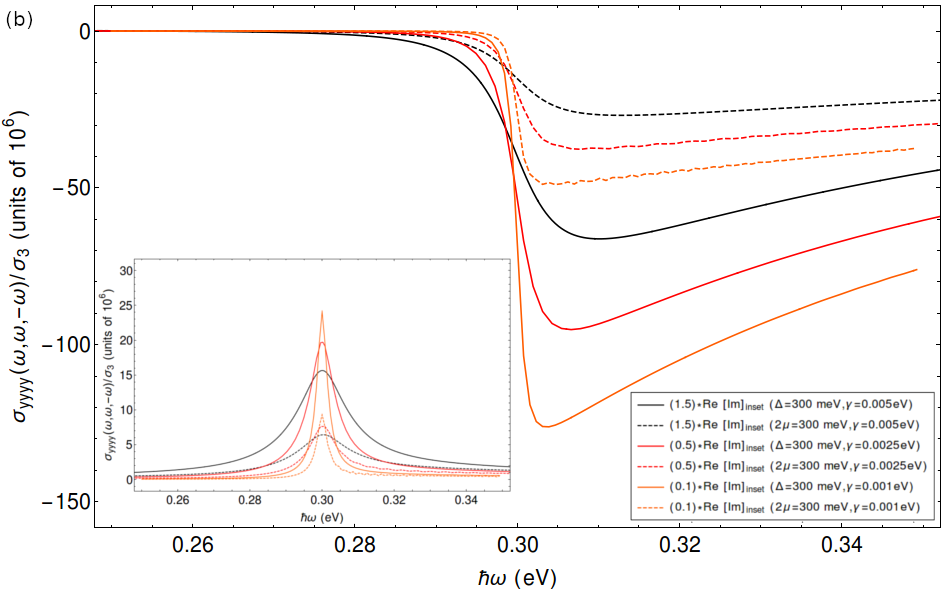}\caption{The real and imaginary (the latter represented in the insets) parts
of the Kerr effect, $\sigma_{yyyy}(\omega,\omega,-\omega)$, in the
GG, $\Delta=300$ meV, and PG, $2\mu=300$ meV, monolayers for frequencies
around $2\hbar\omega\sim\Delta=\,2\mu$ (a), and for frequencies around
$\hbar\omega\sim\Delta=\,2\mu$ (b) for different values of the scattering
parameter: $\gamma=0.005$ eV (black), $\gamma=0.0025$ eV (red) and
$\gamma=0.001$ eV (orange). Note that in (b), the conductivities
for different $\gamma$ have been scaled by different factors: $1.5$
for black, $0.5$ for red and $0.1$ for orange. As before, the vertical
scale in both figures is in units of $10^{6}$. \label{fig:12} }
\end{figure*}

\subsubsection{Optical Kerr Effect (OKE)}

Our final set of results concerns the optical Kerr effect (OKE), once
again calculated for the cases of the GG and PG monolayers of parameters,
$\Delta=2\mu$. Now, unlike the THG \textemdash{} Eq.(\ref{eq:THG_TENSOR_COM})
\textemdash{} not all nonzero components of the conductivity tensor
associated with the OKE can be directly related to the diagonal terms.
To show their differences, we present the $\sigma_{yyyy}(\omega,\omega,-\omega)$
and $\sigma_{yxxy}(\omega,\omega,-\omega)$ components in the low-frequency
portion of the response, i.e. around the one and two photon processes
at the gap (twice the chemical potential for the PG), Figures \ref{fig:fig9}
and \ref{fig:10}(a), as well as the response around the one photon
process at the van Hove singularity, Figure (\ref{fig:10})(b). Figure
\ref{fig:fig9} shows that the two conductivity components of the
GG monolayer have opposite signs \textemdash{} in both the real and
the imaginary part \textemdash{} and that they are similar, but not
exactly equal, in modulus. In the PG, it is only the height of the
features that is different, being less pronouced in the off-diagonal
component. For the two photon processes at the gap (twice the chemical
potential), Figure \ref{fig:10}(a), both the GG and PG conductivities
display sign differences between the two tensor components and the
property observed in the one photon seems to appear in reverse: here
it is in the response of the GG that we see the less pronounced features
for the off-diagonal component; the PG conductivities have opposite
signs and are rather similar, in modulus, across the frequency range
considered. For the high frequency response, i.e. one photon processes
at the van Hove singularity, Figure \ref{fig:10}(b), we see that
the features on both components of the OKE conductivity are essentially
the same, differing only by an overall factor of three. As in the
THG, the only distinction between responses of the GG and PG monolayers
comes from the fact the different energy values for the van Hove singularity:
slightly higher in the GG monolayer, Eq.(\ref{eq:VHS}). 

A second point of interest in the OKE concerns the existence of a
divergence in the real part of its associated conductivity for frequencies
above the one photon absorption at the gap (twice the chemical potential)
in the scatteringless limit \citep{Aversa1995}, that is related to
the acceleration of electron-hole pairs \textemdash{} produced in
one photon absorption processes \textemdash{} by a static, nonlinear,
electric field. This divergence should be present in both the GG and
PG monolayers and was indeed seen in an analytical calculation of
the OKE in the monolayer of plain graphene, in the context of a linearized
band \citep{Cheng2014}. Although we cannot probe this singularity
directly \textemdash{} in the sense that the scattering parameter
is necessarily finite in the numerical calculations \textemdash{}
we find that it is nonetheless clear that such a divergence does exists,
in both the PG and the GG monolayer. Figure (\ref{fig:12}) represents
the real and imaginary parts of the OKE conductivity for frequencies
around the two photon (a) and one photon (b) at the gap (twice the
chemical potential) for different values of the scattering parameter,
$\gamma$. For frequencies, $2\hbar\omega\sim\Delta=2\mu$, Figure
(\ref{fig:12})(a), we can see that a decrease in the value of $\gamma$
is associated with sharper features in a small region around the absorption
threshold, that then tend to merge as one moves to frequencies away
from those around the threshold. This is similar to what we have observed
in Figures \ref{fig:4}(b) and \ref{fig:6}(b) and it is the expected
behavior for features in any given regular conductivity. When we move
to frequencies above the one photon absorption, $\hbar\omega\ge\Delta=2\mu$,
this no longer holds for the real part of the OKE conductivity. It
increases in absolute value as $\gamma$ is reduced with the curves
for different $\gamma$ running parallel to one another. Instead of
a well-localized feature, one can see the appearance of a divergence.

\section{SUMMARY}

In this work we have studied second and third harmonic generation,
the optical rectification and the optical Kerr effect for the gapped
and plain graphene monolayers to a monochromatic pulse by using the
density matrix formalism in the velocity gauge as well as in the length
gauge. Although the topic is not new, this is the first work to present
all tensor components of the nonlinear conductivities of these materials,
in a frequency range that extends beyond the Dirac approximation.
We emphasize that the tensor components considered here are not the
effective tensors of ref.\citep{Hipolito2018}, the use of which,
we think, has not been adequately justified.

To calculate the conductivities in this work, we used the velocity
gauge formalism developed in a previous work \citep{passos} with
an additional point that we presented here: the choice of an adequate
basis \textemdash{} the second Bloch basis \textemdash{} can be used
to reduce covariant derivatives to regular \textbf{k}-space derivatives,
which in turn simplifies the computation of the $h$ coefficients
that are required for the calculation of nonlinear optical responses
in the velocity gauge. We have also shown how this treatment of the
covariant derivative is related to the representation of the position
operator, the choice of which bears an influence in the results.

As for the nonlinear conductivities themselves: for the second harmonic
generation and the optical rectification conductivity at the gap,
the numerical results of the velocity gauge were complemented by analytical,
zero scattering limit, results in the length gauge. From these numerical
results we saw how the interplay between the gap, $\Delta$, and the
scattering rate, $\gamma$, affected the form of the features at low
frequency. For higher frequencies, that is, around the van Hove singularity,
we saw the relation between conductivities of GG monolayers with different
values of the gap as well as a blueshift of the features for increasing
values of $\Delta$. For the third order response, we instead focused
on a comparison between the responses of the gapped graphene and doped
plain graphene monolayer, in the case where the excluded energy region
for interband transitions is the same, i.e., $\Delta=2\mu$. We saw,
in the case of the THG, that the low frequency limit in the two materials
is very different, which can be traced back to the presence of intraband
terms in the response of the doped PG monolayer. For higher frequencies,
the two responses are very much alike, with the exception of the shift
that follows from the different location of the van Hove singularity.
For the OKE, we studied two different components of the conductivity
tensor, for both low and high frequencies, as well as the existence
of a divergence for frequencies above the one photon absorption at
the gap (twice the chemical potential) in the response of both the
PG and the GG monolayers.

The authors acknowledge financing of Funda\c{c}\~{a}o da Ci\^{e}ncia
e Tecnologia, of COMPETE 2020 program in FEDER component (European
Union), through projects POCI-01-0145-FEDER-028887 and UID/FIS/04650/2013.
G. B. V. would like to thank Emilia Ridolfi and V\'{i}tor M. Pereira
of the Centre for Advanced 2D Materials at the National University
of Singapore for useful discussions, Nuno M. R. Peres for his help
reviewing the manuscript and Sim\~{a}o Meneses Jo\~{a}o for providing
Figure \ref{fig:honey}.

\end{document}